# Mining Flipping Correlations from Large Datasets with Taxonomies

Marina Barsky† Sangkyum Kim‡ Tim Weninger‡ Jiawei Han‡

† Univ. of Victoria, BC, Canada, marina_barsky@gmail.com

‡ Univ. of Illinois at Urbana-Champaign, {kim71, weninger1, hanj}@illinois.edu

## ABSTRACT

In this paper we introduce a new type of pattern – a flipping correlation pattern. The flipping patterns are obtained from contrasting the correlations between items at different levels of abstraction. They represent surprising correlations, both positive and negative, which are specific for a given abstraction level, and which "flip" from positive to negative and vice versa when items are generalized to a higher level of abstraction. We design an efficient algorithm for finding flipping correlations, the Flipper algorithm, which outperforms naïve pattern mining methods by several orders of magnitude. We apply Flipper to real-life datasets and show that the discovered patterns are non-redundant, surprising and actionable. Flipper finds strong contrasting correlations in itemsets with low-to-medium support, while existing techniques cannot handle the pattern discovery in this frequency range.

## 1. INTRODUCTION

One of the central tasks in data mining is finding correlations in binary relations. Often, this is formulated as a *market basket* problem [1], in which items occurring together are organized into a set of transactions (market baskets). The central goal of this line of work is to find correlations among items based on their recurrent co-appearances among the set of transactions. Such correlations represent the similarity of the correlated items in respect to their togetherness – *e.g.*, items "bought together", words "used together", genes "mutated together". They present valuable information, and the market basket concept has been successfully applied to various domains such as climatology [21], public health [4], and bioinformatics [11, 24].

Typically, a set of one or more items is called an *itemset*. The number of transactions that contain a particular itemset is referred to as the itemset's *support*. To find if particular items in an itemset are correlated, the support of the itemset must be compared with the support of each individual item in it. This is a way to determine both positive (often appear together) and negative (rarely appear together) correlations. Note that mining positive correlations is *not* equivalent to mining frequent itemsets. An itemset can be frequent without positive correlation between items, and very strong positive correlations can be discovered in itemsets with low-to-medium support.

While very frequent itemsets can be efficiently mined due to the anti-monotonicity of support, an efficient algorithm for computing positively or negatively correlated items with low support is a challenge because most useful correlation measures are neither monotonic, nor anti-monotonic. This is especially true if both positive and negative correlations are of interest: negative correlations imply that we need to deal with itemsets with low support. In transactional databases where the number of distinct items is large such computation remains infeasible [19]. In this work we forsake the goal of mining all positive and negative correlations in favor of mining a new type of correlation described below.

In many cases, the transactional data about the relative behavior of the items is accompanied by an additional information, based on intrinsic properties of these items. Each item may be described with differing amounts of detail at different levels of abstraction. For example, whole milk at a higher level is simply milk, and bagels can be generalized as bread. At the next level, both milk and bread can be generalized as grocery products, and so on. Each higher level of abstraction encompasses a group of several items and hence this information can be modeled as a taxonomy tree (*is a* hierarchy). The leaves of a taxonomy tree (or simply taxonomy) represent items at the lowest level of abstraction. Each internal node is by itself an object, or an item, but at a higher abstraction level. The taxonomy tree is generated manually or automatically based on some notion of similarity between objects.

Our goal is to explore correlation differences across abstraction levels in the taxonomy. Specifically, we identify a particular type of correlation called *flipping correlation*, in which the correlation value at one level of abstraction is in contrast with higher-level correlations. That is, the correlation "flips" from positive to negative and vice versa. Furthermore, in order to avoid the significant costs involved with frequent itemset mining, we mine flipping patterns directly, based on correlation values.

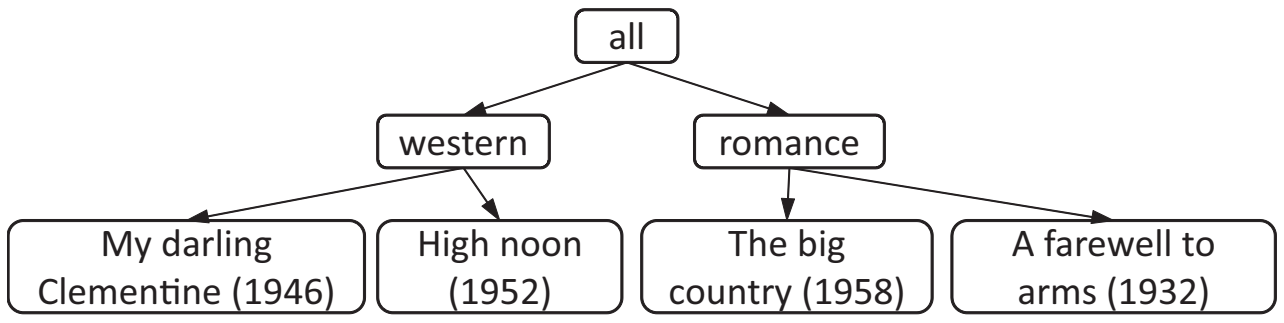

**Figure 1: Fragment of taxonomy tree for Movies dataset: the correlations can be computed between specific movies (items), or between general genres (their generalizations), if each movie is substituted by its genre in the transaction.**

As a motivating example, consider the following correlations extracted from the MovieLens dataset[1], which contains movie rankings, and a hierarchy of movie genres.

Example 1. *To apply the market basket concept to movie rankings, we model each user as a single transaction. Each transaction contains all movies which this user ranked highly (at least 4 out of 5), giving us each user's favorite movies. We can easily find correlations between movies, that is, which sets of movies are almost always favored together. If we replace each movie by its higher-level abstraction, movie genre, then we can find that users who like action movies also like adventure movies, but people who like romance movies rarely also like westerns (negative correlation). However, for the negatively correlated romance and western genres, we found two movies, shown in Figure 2(a), which are positively correlated:* The Big Country (1958) *and* High Noon (1952). *Thus, the positive correlation between these two movies is in contrast with negative correlation between their higher-level concepts.*

*This raises several questions: what is special about these two movies? Why do they stand out from other movies of the same genres, which tend not to be favored by the same users? Here are three potential explanations:*

*(1) These are very good movies and the romance-lovers who do not generally watch westerns, make an exception for* High Noon.

*(2) One of the movies was assigned to a wrong genre.*

*(3) Despite the fact that these movies belong to different genres, they share something which is common to both of them, and thus they present a link between two higher-level abstractions.*

This is an example of a correlation which flips from negative to positive when moving down the branches of the taxonomy tree into a more detailed level of abstraction. It demonstrates the surprising connection between the objects, and sets these objects apart from their siblings, which do not have contrasting behavior towards their generalizations.

Correlations that flip from positive to negative can also be valuable, as can be seen with data from the Groceries dataset [5] in Figure 2(b). Here a negative correlation between eggs and fish is highlighted by the fact that their generalizations are highly positively correlated.

The novelty of the flipping correlation concept is in its contrasting nature. Previously, the taxonomy information was used to characterize only *positive* correlations between items (in the form of association rules with significantly different confidence levels [17]) or in order to rank surprisingness of frequent itemsets based on the distance between items in the taxonomy tree [6]. Unlike previous studies, we are interested in patterns which present sharp flips between positive and negative correlations.

[1]http://www.grouplens.org/node/12

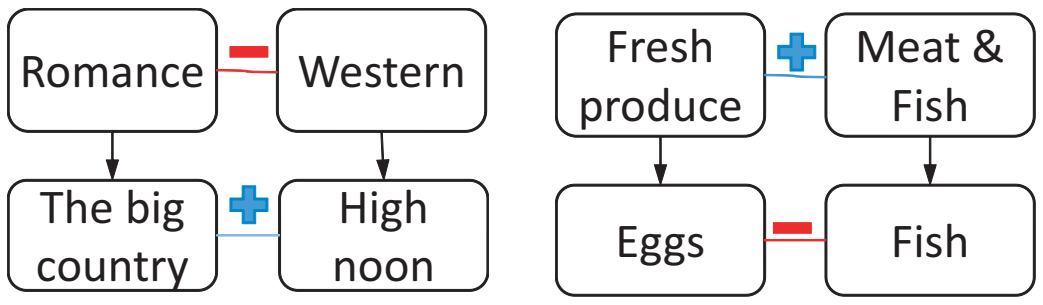

(a) Movies Dataset. (b) Groceries Dataset.

**Figure 2: Sample flipping correlations.**

Thus, in this work we address the problem of efficiently computing all flipping correlations. In previous works, pattern pruning or deduplication was mainly performed as a post-processing step, after first computing all frequent itemsets. Because computing frequent itemsets can be a significant computational challenge, we develop a new method for the efficient computation of flipping correlations *directly*, by proposing novel pruning techniques based on new properties of the selected correlation measures. Moreover, instead of computing all positive and negative correlations and choosing the flipping among them, we push the contrast ("flipping") constraint into the mining process, and use it to improve the efficiency of our algorithm.

We are solving the following problem: how to find flipping correlations *without* generating all frequent itemsets. This task is challenging due to: (1) flipping patterns contain negative correlations which by definition are in itemsets with very low support, (2) computing all frequent itemsets with very low support is computationally prohibitive, and (3) most of the correlation measures which can be applied to large datasets possess neither monotonicity nor anti-monotonicity properties, and as such cannot be straightforwardly used for pruning purposes. We solve this challenging problem by developing new efficient pruning methods.

In this work we make the following contributions:

- We introduce *flipping correlation patterns* and formalize the problem of *flipping correlation mining* (Section 2).
- We present new properties of selected correlation measures, which allow direct pruning based on correlation values instead of support (Section 4.1). Based on these properties, we design an efficient solution for finding all flipping correlations in transactional databases supplied with taxonomy hierarchies (Section 4). The proposed solution is applicable to any known correlation measure that possesses null(transaction)-invariance, including measures that have never been used for pruning before, due to the lack of anti-monotonicity.
- We evaluate the efficiency of new pruning techniques on a variety of synthetic and real datasets and demonstrate examples of non-trivial flipping correlations which could not be discovered using previous techniques (Section 5).

The rest of the paper is organized as follows. In Section 2 we formally define the problem of flipping pattern mining and discuss the selection of a suitable correlation measure. In Section 3 we present and prove new useful properties of these correlation measures. In Section 4 we present Flipper, an algorithm for mining flipping correlation patterns.

The experimental evaluation of FLIPPER is described in Section 5. The previous work and how it differs from our study is discussed in Section 6. Section 7 concludes this study and offers avenues for future research.

# 2. PRELIMINARIES

We start by defining the problem of mining contrasting level-specific correlations, which we call *flipping correlations*. The first step is to choose the measure which is most suitable for our problem, and the second step is to define positive and negative correlations based on this measure.

## 2.1 Correlation measure

Various measures were proposed for assessing the degree of the correlation. A comprehensive comparison of 21 different correlation measures was given by Tan et al. [18]. The popular correlation measure *Lift* accompanied by a $\chi^2$ test for statistical significance [3] is based on *support expectation*. Other expectation-based measures are $\phi$ [2] and the deviation from the expected [23]. To compute *Lift*, the items in the transactional database are treated as binary variables, and the expected support for an itemset containing both $A$ and $B$ is computed as $E(sup(AB)) = \frac{sup(A)}{N} \cdot \frac{sup(B)}{N} \cdot N$, where $N$ is the total number of transactions. If $sup(AB) > E(sup(AB))$, then items $A$ and $B$ are *positively correlated*. Similarly, if $sup(AB) < E(sup(AB))$, then items $A$ and $B$ are *negatively correlated*. The degree of the positive or negative correlation is measured by the degree of the deviation of the real support from the expected one.

Unfortunately, the expectation-based measures are unreliable when used for assessing the degree of the correlation in large transactional databases [22]. As an illustration consider the following example.

EXAMPLE 2. *Consider two sample databases ($DB_1$ and $DB_2$) shown in Table 1. These databases differ only by N, but not by the number of transactions containing itemsets A, B, AB and C, D, CD. One can see that the relationship between items A and B and that between C and D can be classified either as a positive or as a negative correlation, solely depending on the total number of transactions N, instead of reflecting the true relationships between the items. For example, C and D, though intuitively a clearly* negative *correlation, is judged as* positive *by the expectation-based correlation measure in dataset $DB_1$. Thus, expectation-based correlation measures are unstable and cannot be used to produce meaningful positive and negative correlations.*

Since in large databases the number of transactions which contain particular item is much smaller than the total number of transactions $N$ (small-probability event), the expected value for support for both itemsets $AB$ and $CD$ will be extremely low in Database $DB_1$. Then even very small actual support will be greater than the expected, and both correlations will be classified as positive.

The degree of the expectation-based correlation is highly influenced by the number of *null transactions* [20, 22], *i.e.*, transactions which do not contain items whose correlation has been measured. Hence, such measures are not suitable for the study of correlations in large datasets, where the number of null transactions could be large and unstable.

For our problem of contrasting positive and negative correlations, it is crucial to adopt a reliable correlation measure that is unconcerned with the number of null-transactions present in the database. These measures are called *null (transaction) - invariant* [20]. The main property of a null-invariant measure is its independence of the total number of transactions $N$.

According to the study of Wu et al. [22], all five known null-invariant measures can be viewed as a generalized mean of conditional probabilities. The conditional probabilities represent how many transactions containing item $A_i$ also contain the rest of the items, and an average over these probabilities assesses the degree of the mutual connection between items in the itemset. Thus, the degree of this connection is based solely on the number of relevant transactions, i.e. the transactions that contain at least one item in the itemset to be evaluated. The five measures are summarized in Table 2. The ordering of the measures for the same conditional probabilities follows from the nature of a mean which they represent:

$$\begin{array}{ccccc} Coherence(A_1A_2) & \leq & Cosine(A_1A_2) & \leq & Kulc(A_1A_2) \\ \text{harmonic mean} & \leq & \text{geometric mean} & \leq & \text{arithmetic mean} \end{array}$$

Depending on the support counts of single items, the measures produce different results: if $sup(A_1)$ is much larger than $sup(A_2)$, the $Coherence$ value of such an itemset tends to be small, no matter how strong is the relationships between the items, while $Kulc$ value will be large if such strong relationship exists. Hence, the different correlation measures are incomparable, and in order to handle both positive and negative correlations, it is bettter to use the same consistent correlation measure throughout the entire mining process. The discussion about the choice of the most appropriate correlation measure can be found in work of Wu et al. [22]. Our method can be performed using any null-invariant measure. As an illustration, we use $Kulczynsky$ (denoted as $Kulc$) for our experiments. $Kulc$ is a relaxed measure and it handles unbalanced itemsets better than $Coherence$ and $Cosine$ [22]. Similar to $Cosine$ and $Max\ confidence$, mining correlations using $Kulc$ represents also a computationally challenging case, since these three measures are not anti-monotonic.

Let $Corr$ be one of the null-invariant correlation measures from Table 2. We formally define positive and negative correlations as follows. Recall that an itemset is *frequent* if its support is not less than a minimum support threshold $\theta$ predefined by a domain expert. For our problem, the minimum support threshold $\theta$ can be arbitrarily low.

*Definition 1.* **Null-invariant correlations.** Items in a $k$-itemset $A = \{a_1, \dots, a_k\}$ are *positively correlated* with a correlation measure $Corr$ if $A$ is *frequent* and $Corr(A) \geq \gamma$ for a positive correlation threshold $\gamma$. Items in $A$ are *negatively correlated* if $A$ is *frequent* and $Corr(A) \leq \varepsilon$ for a negative correlation threshold $\varepsilon$.

In our experiments we use $Corr(A) = Kulc(A)$. However, in the next section we show that all the techniques developed here are applicable to any known null-invariant measure, and the efficiency of our new algorithm is not influenced by the concrete choice of the correlation measure. Based on Definition 1 we formally define the problem of flipping correlations mining.

[1]This re-definition of $Coherence$ which preserves the features and the ordering of the original $Coherence$ measure [22].

**Table 1: Examples of the expectation-based correlation.**

| $Kulc(A,B)=0.40$ | | | | | | |
|---|---|---|---|---|---|---|
| | sup($A$) | sup($B$) | sup($AB$) | Total $N$ | $E(sup(AB))$ | Correlation |
| $DB_1$ | 1,000 | 1,000 | 400 | 20,000 | 50 | *positive* |
| $DB_2$ | 1,000 | 1,000 | 400 | 2,000 | 500 | *negative* |

| $Kulc(C,D)=0.02$ | | | | | | |
|---|---|---|---|---|---|---|
| | sup($C$) | sup($D$) | sup($CD$) | Total $N$ | $E(sup(CD))$ | Correlation |
| $DB_1$ | 200 | 200 | 4 | 20,000 | 2 | *positive* |
| $DB_2$ | 200 | 200 | 4 | 2,000 | 20 | *negative* |

| Name | $Corr(A_1A_2)$ (2 items) | $Corr(A_1\dots A_k)$ ($k$ items) | Description |
|---|---|---|---|
| *All Confidence* | $min(P(A_1\|A_2),P(A_2\|A_1))$ | $min_{i=1}^{k}(P((A_1\dots A_k)\|A_i))$ | minimum |
| *Coherence*[1] | $2\times(P(A_1\|A_2)^{-1}+P(A_2\|A_1)^{-1})^{-1}$ | $k\times(\sum_{i=1}^{k}P((A_1\dots A_k)\|A_i)^{-1})^{-1}$ | harmonic mean |
| *Cosine* | $\sqrt{P(A_1\|A_2)P(A_2\|A_1)}$ | $\sqrt[k]{\prod_{i=1}^{k}P((A_1\dots A_k)\|A_i)}$ | geometric mean |
| *Kulczynsky* | $(P(A_1\|A_2)+P(A_2\|A_1))/2$ | $\sum_{i=1}^{k}P((A_1\dots A_k)\|A_i))/k$ | arithmetic mean |
| *Max Confidence* | $max(P(A_1\|A_2),P(A_2\|A_1))$ | $max_{i=1}^{k}(P((A_1\dots A_k)\|A_i))$ | maximum |

**Table 2: Definitions of five null-invariant correlation measures.**

## 2.2 Flipping pattern

Let $\mathcal{I}$ be a set of items, and let $\mathcal{T}$ and $\mathcal{D}$ be two independent sources of the information about these items. The *taxonomy tree* $\mathcal{T}$ represents the mapping of the items into several levels of abstraction. Each internal node of the taxonomy tree represents a higher-level abstraction for a group of items, and is itself an item. The leaves in $\mathcal{T}$ represent the most specific items, and internal nodes represent more general items. The root of $\mathcal{T}$ represents *all* items in $\mathcal{I}$, and is considered to be at abstraction level 0. Since there is only one node at level 0, we cannot compute correlation for a single item, and we exclude the root node from further consideration. Let *height* $H$ of the taxonomy tree $\mathcal{T}$ be the number of nodes from the top level 1 to the deepest leaf. Then, there are $H$ different abstraction levels in the tree, and each node belongs to some level.

While $\mathcal{T}$ summarizes the intrinsic similarity relationships between items, an additional information about the relative behavior of the same items is presented as a set $\mathcal{D}$ of observations, or *transactions*. This is the source of the information about the correlation of different items.

Recall that any combination of $k$ unique items from $\mathcal{I}$ forms a $k$-*itemset*. The *support* of itemset $A=\{a_1,\dots a_k\}$, $sup(A)$, is the number of transactions containing all items from $A$. In terms of support we measure the correlation between items in $A$ as:

$$Kulc(A)=\frac{1}{k}\sum_{i=1}^{k}\frac{sup(A)}{sup(a_i)}. \quad (1)$$

According to Definition 1, the correlation between items in $A$ is positive if $Corr(A)$ is greater than a user-specified *positive threshold* $\gamma$, and it is negative if $Corr(A)$ is less than a user-specified *negative threshold* $\varepsilon$. If none of these conditions holds, the items in $A$ are considered non-correlated, and not interesting. For convenience, we call the itemset where the items are positively correlated a *positive* itemset, and where the items are negatively correlated a *negative* itemset.

The correlation can be computed between different nodes of $\mathcal{T}$ at the same level of the hierarchy, if we replace the items in transactions by their higher-level generalizations. An $(h,k)$-*itemset* ($1\le h\le|\mathcal{I}|$, $1\le h\le H$) is defined as a set of $k$ items from $\mathcal{I}$, replaced by their corresponding generalizations from the level $h$ of the taxonomy tree.

The goal is to find all positive and negative correlations between the nodes of taxonomy tree $\mathcal{T}$ *at the same level of abstraction*. We are interested only in the level-specific correlations of a contrasting nature, *i.e.*, if the correlation between nodes is positive, then the correlation between their minimal generalizations is negative and vice versa. We say that the correlation *flips* from level to level.

*Definition 2.* **Flipping pattern**. A $k$-itemset $A$ represents a *flipping (correlation) pattern* if all $(h,k)$-itemsets, obtained by replacement of items in $A$ with their corresponding minimal generalizations, have flipping correlation labels. In other words, if an $(h,k)$-itemset is positive, then an $(h+1,k)$-itemset is negative, and vice versa.

Since the goal is to find correlations between *different* items at each level of the hierarchy, all items in a flipping correlation pattern are descendants of different nodes at hierarchy level 1.

By definition 2, a flipping correlation pattern is an itemset which has flipping correlations *across the entire height* $H$ of the taxonomy tree. Note that this definition is general enough to satisfy any possible user query for contrasting level-specific correlations: if the level-specific correlations are required for a specific subset of all levels, all that needs to be changed is the input to the algorithm, which would be a truncated taxonomy tree containing these specific levels of interest.

Since we target the correlations at the same level of abstraction, in case that the depth of some item-leaf node $v_i$ is less than $H$, it is the user's responsibility to define missing corresponding generalizations of $v_i$. In Figure 3 we show some possible methods of dealing with such situations. In our experiments we rebalanced the tree by adding additional copies of $v_i$ as its descendants up to depth $H$ (Figure 3 [B]).

To demonstrate that replacing items by their generelizations may indeed drastically change the degree of the correlation consider the following example.

EXAMPLE 3. *In Figure 4, we show a toy example of 10 transactions and a taxonomy tree of the corresponding items. The input database has 8 different items from 2 different categories a and b. Items in each transaction can be substituted by their generalizations. Given positive threshold* $\gamma=0.6$ *and negative threshold* $\varepsilon=0.35$*, we find that there is only one itemset,* $\{a_{11},b_{11}\}$*, which is a flipping correlation pattern (Figure 5).* ∎

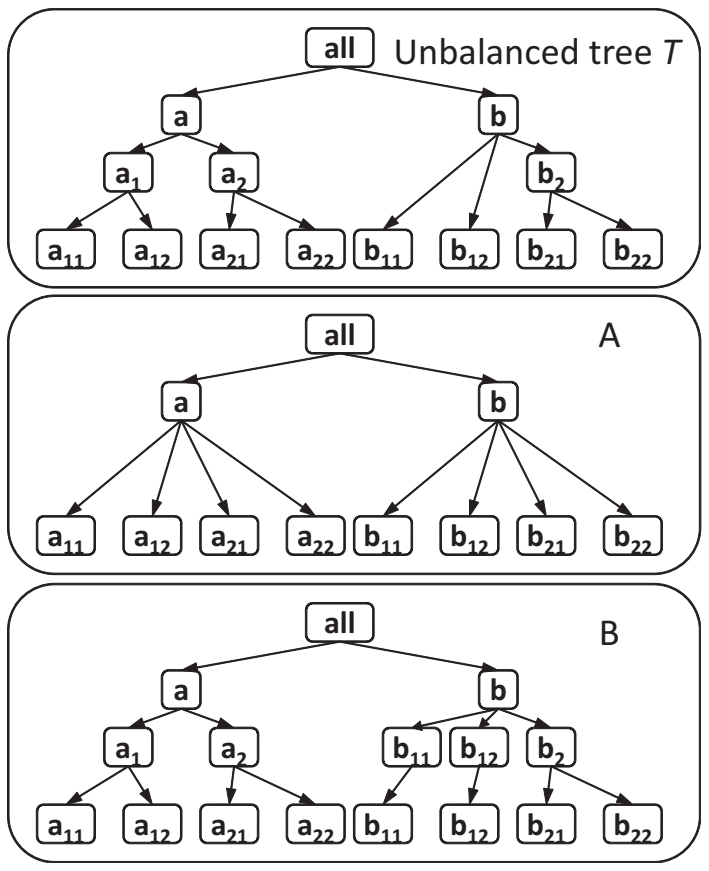

**Figure 3: Variants of re-balancing the levels of taxonomy tree: [A] truncate tree by leaving only consistent levels; [B] consider the copies of leaf nodes as their generalizations.**

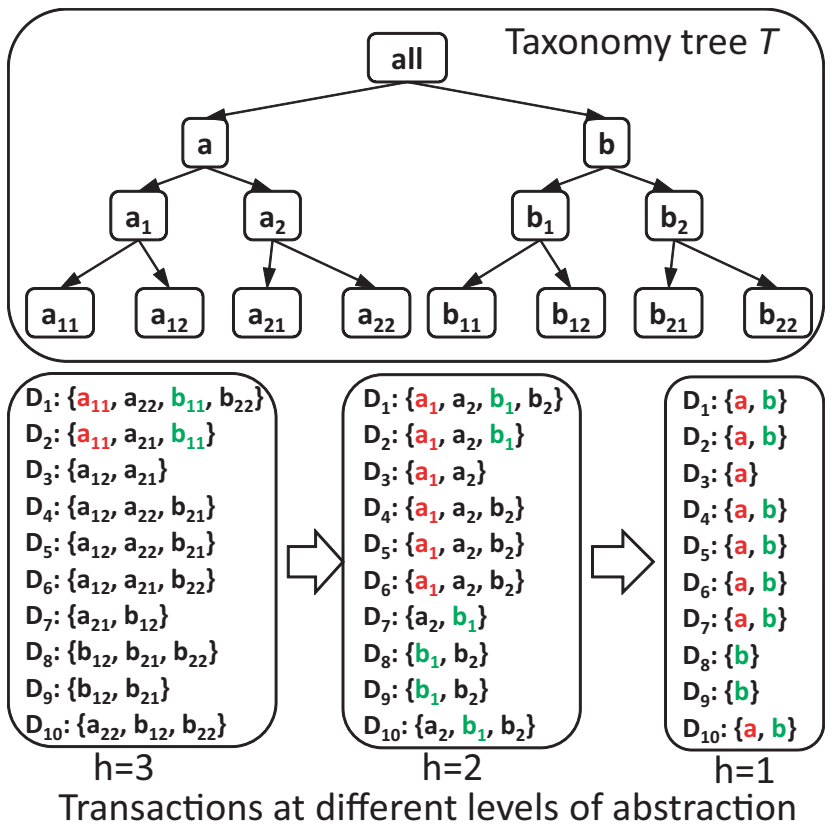

**Figure 4: A toy example of a taxonomy and a database of 10 transactions.**

For mining frequent itemsets at different levels of $\mathcal{T}$, it is useful to define different minimum supports for each level, because items at lower levels of abstraction are unlikely to occur as frequently as those at higher levels. If we use low support for the highest hierarchy level, we will end up with too many branching itemsets. We assume that a set $\{\theta_1, \ldots, \theta_H\}$ of non-increasing support thresholds is provided as an input to our algorithm. Though we use the support-based pruning in our computation, we do not mainly rely on it. Hence, the support thresholds can be set arbitrarily low.

With the above definitions, **the flipping pattern mining problem** can be stated as follows:
*Input:* A set of transactions $\mathcal{D}$, a taxonomy tree $\mathcal{T}$, a set of thresholds: $\gamma$, $\varepsilon$, and minimum support $\theta_h$ for $1 \leq h \leq H$.
*Output:* All flipping correlations satisfying the thresholds.

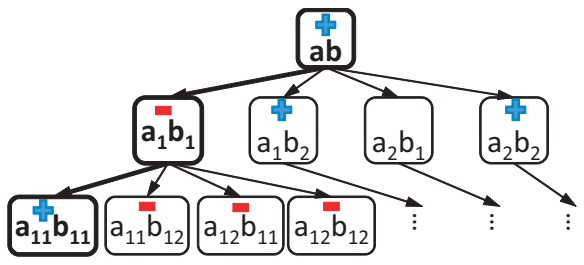

**Figure 5: An example of a flipping pattern from dataset in Figure 4.**

# 3. NEW PROPERTIES

Before presenting the solution to the problem of flipping correlation mining, we describe and prove useful mathematical properties common to all null-invariant correlation measures. These properties constitute the basis of an efficient flipping pattern mining algorithm, presented in Section 4.

The anti-monotonicity of *support* allows to systematically control the exponential growth of candidate itemsets. The superset of non-frequent itemset $A$ cannot be frequent: if we are adding one more item to $A$, the new item combination cannot occur more often than $A$. In contrast, adding an additional item to $A$ may increase or decrease the value of the *correlation* between items in a new itemset: most of null-invariant correlation measures, being generalized means, are not anti-monotonic. The lack of the anti-monotonicity poses a significant challenge, if the efficiency of support-based candidate pruning needs to be enhanced with the correlation-based pruning. The two properties of correlation measures presented below are intended to overcome this limitation.

## 3.1 Correlation upper bound

The following theorem proves an upper bound of a correlation value of a superset in terms of correlation values of its subsets. It reflects an intuitive observation that correlation of a superset cannot be positive if all its subsets are non-positive.

THEOREM 1. **Correlation upper bound**
*For $k$-itemset $A$ and a set $S$ of all $A$'s ($k$-1)-subitemsets, $Corr(A) \leq max_{B \in S}(Corr(B))$.*

PROOF. Let $A = \{a_1, \cdots, a_k\}$, and $B^i = A - \{a_i\}$ be a subset of $A$ which contains all elements of $A$ except $a_i$, for $i = 1, \cdots, k$.

Because $sup(B^i) \geq sup(A)$ (anti-monotonic), $P(A|a_j) \leq P(B^i|a_j)$ for any $1 \leq j \leq k$. Hence, the theorem trivially holds for *All confidence* and for *Max confidence* (minimum and maximum of conditional probabilities respectively). The proof for *Coherence*, which essentially is $sup(A)$ divided by support of all transactions containing any item from $A$ (intersection over union), is straightforward: the numerator in formula of *Coherence* is non-increasing, and the denominator is non-decreasing, while adding one more item to itemset $B^i$. Only *Kulc* and *Cosine* require a special treatement.

**Proof for** *Kulc*. The arithmetic mean of *Kulc* values of all $B^i$'s is:

$$\frac{1}{k}\sum_{i=1}^{k}(Kulc(B^i)) \quad = \quad \frac{sup(B^i)}{k}(\frac{1}{sup(a_1)} + \ldots + \frac{1}{sup(a_k)})$$

Here each element $\frac{1}{sup(a_i)}$ appears in the sum ($k$-1) times. Since $sup(B^i) \geq sup(A)$, replacing $sup(B^i)$ with $sup(A)$ gives the following inequality:

$$\frac{1}{k}\sum_{i=1}^{k}(Kulc(B^i)) \geq \frac{sup(A)}{k}(\frac{1}{sup(a_1)} + \ldots + \frac{1}{sup(a_k)}) = Kulc(A)$$

Since the maximum is not smaller than the arithmetic mean, we have proven that $Kulc(A) \leq max_{B \in S}(Kulc(B))$

**Proof for** *Cosine*. The geometric mean of *Cosine* values for $B_i$ is:

$$\sqrt[k]{\prod_i^k Cosine(B^i)}$$

$$= \sqrt[k]{\frac{sup(B^1)}{\sqrt[k-1]{sup(a_2) \times \ldots sup(a_k)}} \times \ldots \frac{sup(B^k)}{\sqrt[k-1]{sup(a_1) \times \ldots sup(a_{k-1})}}}$$

Each element $\frac{1}{\sqrt[k-1]{sup(a_i)}}$ is multiplied ($k$-1) times. Since

$sup(B^i) \geq sup(A)$, replacing $sup(B^i)$ with $sup(A)$ again gives:

$$\begin{aligned} \sqrt[k]{\textstyle\prod_i^k Cosine(B^i)} &\geq \frac{sup(A)}{\sqrt[k]{sup(a_1)} \times \ldots \times \sqrt[k]{sup(a_k)}} \\ &= Cosine(A) \end{aligned}$$

Since the maximum is not smaller than the geometric mean, we have proven that $Cosine(A) \leq max_{B \in S}(Cosine(B))$.

This completes the proof of Theorem 1 for all five null-invariant correlation measures. ∎

The following corollary follows directly from Theorem 1:

Corollary 1. *If all (k-1)-sub-itemsets of a k-itemset A are non-positive, then A cannot be positive.*

## 3.2 Itemsets with a special single item

Recall that if we use the measure *Corr* which is not anti-monotonic, we cannot imply that a superset of some itemset $A$ is non-positive, even if $Corr(A) < \gamma$. However, in the following, we claim that for an item $a$ with special properties, knowing correlation values of all ($k$-1)-itemsets containing this item $a$ allows to evaluate all $k$-itemsets containing $a$.

Theorem 2. *For k-itemset $A = \{a_1, \ldots, a_k\}$, and all its ($k$-1) subsets of size ($k$-1), which share the same single item a, if (1) the correlation values for all these subsets are below $\gamma$ and (2) the support of at least one item $a_i \neq a$ in $A$ is greater than or equal to $sup(a)$, then correlation between items in $A$ is below $\gamma$.*

Proof. Theorem trivially holds for *Coherence* and *All confidence*, which are anti-monotonic. For anti-monotonic measures, if correlation values for all itemsets containing item $a$ are below threshold, then none of their supersets can be positive. Hence, Theorem 2 holds without condition (2). The proofs for *Kulc*, *Cosine* and *Max Confidence* are presented below.

Assume that the first item of $A$ $a_1 = a$ and the last item $a_k$ has the largest support among all single items in $A$, without loss of generality.

**Proof for** *Kulc*. By simple algebra we can show that:

$$\frac{1}{k-1} \sum_{i=1}^{k-1} \frac{1}{sup(a_i)} \geq \frac{1}{k} \sum_{i=1}^{k} \frac{1}{sup(a_i)}.$$

Then,

$$\begin{aligned} Kulc(A) &= \frac{sup(A)}{k} \sum_{i=1}^{k} \frac{1}{sup(a_i)} \\ &\leq \frac{sup(A)}{k-1} \sum_{i=1}^{k-1} \frac{1}{sup(a_i)} \\ &\leq \frac{sup(A - \{a_k\})}{k-1} \sum_{i=1}^{k-1} \frac{1}{sup(a_i)} \\ &= Kulc(A - \{a_k\}) \\ &< \gamma, \end{aligned}$$

where $A - \{a_k\}$ represents a ($k$-1)-subset of $A$ which does not contain item $a_k$, and, by condition (1), its correlation is below the positive threshold as for any of the ($k$-1)-itemsets containing $a_1$. This proves theorem 2 for *Kulc*.

| | | k-itemsets | | | |
|---|---|---|---|---|---|
| | | k=2 | k=3 | … | k=K |
| Hierarchy | h=1 | | | | |
| | h=2 | | $Q_{2,3}$ | | |
| | ⋮ | | | | |
| | h=H | | | | $Q_{H,K}$ |

**Figure 6: Search space for flipping correlations.**

**Proof for** *Cosine*. By similar simple algebra, we can show that

$$\sqrt[k-1]{sup(a_1) \times \cdots \times sup(a_{k-1})} \leq \sqrt[k]{sup(a_1) \times \cdots \times sup(a_k)}.$$

Then

$$\begin{aligned} Cosine(A) &= \frac{sup(A)}{\sqrt[k]{sup(a_1) \times \cdots \times sup(a_{k-1}) \times sup(a_k)}} \\ &\leq \frac{sup(A)}{\sqrt[k-1]{sup(a_1) \times \cdots \times sup(a_{k-1})}} \\ &\leq \frac{sup(A - \{a_k\})}{\sqrt[k-1]{sup(a_1) \times \cdots \times sup(a_{k-1})}} \\ &\leq Cosine(A - \{a_k\}) \\ &< \gamma, \end{aligned}$$

This completes the proof for *Cosine*.

The **proof for** *Max confidence* is straightforward: if all ($k$-1)-itemsets which contain item $a_1$ are non-positive, we can always represent a superset of any of them as adding one more item $a_k$, with support which is maximum between all supports of $a_i$. By condition (2) we know that such item $a_k$ exists and is different from $a_1$. However, the conditional probability we are adding as an argument to the *max* function has numerator which is non-increasing ($sup(A)$), and denominator which is the greatest among all supports considered for a ($k$-1)-subitemset. Hence, we cannot create a positively correlated itemset by adding this new item.

We have proven that Theorem 2 holds for all five null-invariant correlation measures. ∎

The following corollary follows directly from Theorem 2.

Corollary 2. *If the maximum Corr for all k-itemsets containing item a is less than $\gamma$, and item a has the smallest support between single items existing in the database, then Corr of all $k'$-itemsets containing a is less than $\gamma$ for $k' \geq k$.*

Proof. Each $(k+1)$-itemset $A'$ which contains $a$ can be thought of as an extension of some $k$-itemset containing $a$ with an item $a_{k+1}$, which has the largest support among all the items in $A'$ (since we know that support of $a$ is not the largest). Then, by Theorem 2, $Corr(A') < \gamma$. Since all $k$-itemsets containing item $a$ have *Corr* value less than $\gamma$, all $(k+1)$-itemsets containing $a$ have *Corr* value less than $\gamma$. Iteratively applying Theorem 2, now to extension of $(k+1)$-itemsets into $(k+2)$-itemsets, containing $a$, we conclude that none of the $k'$-itemsets containing $a$ is positive, for $k' \geq k$ ∎

# 4. FLIPPER ALGORITHM

This section describes our solution to the flipping pattern mining problem. The goal is to extract all sequences of $(h, k)$-itemsets such that (1) each itemset is either positive or negative, (2) any two consecutive ($h$-1, $k$)- and $(h, k)$-itemsets have opposite correlation signs (flip), and (3) the flipping sequence continues unbroken from the top to the bottom of the hierarchy. Our search space includes all $(h, k)$-itemsets for $1 \leq h \leq H$ and $2 \leq k \leq K$, where $K$ is the number of distinct items in the largest possible itemset. We model the search space as a two-dimensional table $\mathcal{M}$ presented in Figure 6. Each cell $Q_{h,k}$ of this table contains $k$-itemsets, where items are substituted by their generalizations from hierarchy level $h$.

## 4.1 General framework

The number of possible itemsets in each cell of table $\mathcal{M}$ grows exponentially from left to right and from the top to the bottom. This suggests the top-down and left-to-right directions for exploring table $\mathcal{M}$. For convenience, we call the extension of $(h, k)$-itemsets into $(h + 1, k)$-itemsets the *vertical* pattern growth, and the extension of $(h, k)$-itemsets into $(h, k + 1)$-itemsets the *horizontal* pattern growth.

By Definition 2 all items in a flipping pattern are descendants of the generalizations at level 1 of the taxonomy tree, and hence the maximum number of columns in $\mathcal{M}$ is bounded by the total number of different nodes at level 1 of the taxonomy tree, or by the maximum number of distinct items in the same transaction (max transaction width), whichever is less.

For each cell, we first generate the set of candidates, based on the information from the previously computed cells to the left and above it, and then we perform the support counting for these candidates, by checking each transaction for the corresponding item combination. This framework represents a two-dimensional modification of the level-wise frequent pattern mining known as the *Apriori* algorithm. The efficiency of such a level-wise processing depends on the efficiency of *pruning*: some itemsets in $Q_{h,k}$ can be excluded from the set of candidates, if we can infer – from already computed values in cells $Q_{h\text{-}1,k}$ and $Q_{h,k-1}$ – that they cannot be a part of a flipping pattern. The pattern growth terminates when for some cell the set of the candidates is empty.

In the following, we describe how we modify this basic framework using the definition of a flipping pattern and the correlation properties presented in Section 3. These optimizations lead to a significant reduction of the number of candidates in each cell and to the early termination of the pattern growth across both dimensions, which accounts for the high efficiency of our mining algorithm.

## 4.2 Basic pruning

### 4.2.1 Pruning by support

Since by Definition 1 positive and negative correlations are computed only between items in *frequent* itemsets, the first pruning and stopping criteria is pruning by support. If itemset $A$ is non-frequent it is not extended both horizontally and vertically, and the combination of items in $A$ is excluded from further consideration. $A$ is not extended vertically, since we want for all itemsets in a flipping sequence to be either positive or negative, and $A$ is neither, hence it breaks the sequence of flipping correlations. There is no need to extend $A$ horizontally as well, since for the same support threshold $\theta_h$ no supersets of $A$ are frequent, and hence they break the flipping sequence also in subsequent columns of $\mathcal{M}$. Normally, this anti-monotone property of support accounts for an efficient pruning. However, in our case, the arbitrarily low support thresholds have an adversary effect producing an exponentially large number of candidates, which cannot be counted by a single database scan. The pattern growth terminates very late, and the pruning-by-support has a limited value for flipping correlation mining.

### 4.2.2 Pruning non-flipping itemsets

Definition 2 of flipping patterns requires that any two consecutive itemsets $A_p$ (parent) and $A_c$ (child) have opposite correlation signs. Thus, if both $A_p$ and $A_c$ are non-positive (or both are non-negative), they break a flipping sequence and do not need to be extended *vertically*. However, a superset of $A_c$ still can be a part of a flipping pattern, since we cannot predict the correlation value of each superset of $A_c$ based only on the correlation of items in $A_c$. Thus we need to count all supersets of $A_p$ and $A_c$ till the end of the corresponding rows in table $\mathcal{M}$. Hence, we cannot prune supersets of $A_c$ from the candidates until we finish processing two consecutive rows. This suggests the row-by-row order of processing, since the combination of items in $A_c$ should be kept anyway to generate corresponding candidate supersets. When at least 2 rows are completed, all such non-flipping itemsets are excluded from further consideration. Since the final pruning of non-flipping itemsets is performed after the entire row has been processed, this pruning has a limited efficiency as well.

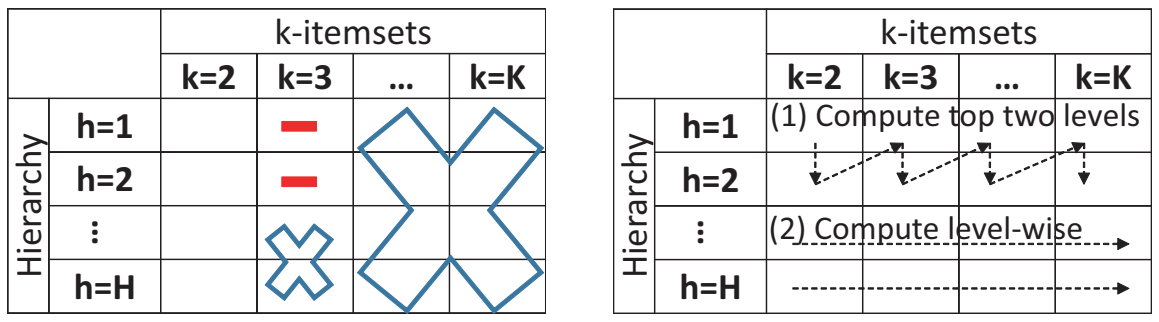

(a) (b)

**Figure 7: (a) Termination of pattern growth if all correlations in two vertically consecutive cells are negative; and (b) Order of processing for the early termination of pattern growth.**

## 4.3 Advanced pruning

### 4.3.1 Early termination of the pattern growth

Theorem 1 presents an upper bound for a correlation value of $k$-itemset $A$ in terms of its sub-itemsets. Corrolary 1 suggests that once correlation values of all ($k$-1)-subitemsets of $A$ fall below positive correlation threshold $\gamma$, we conclude that $Corr(A) < \gamma$. However, this does not imply that all supersets of $A$ are non-positive: there might be a superset of $A$, $A' = \{a_1, \ldots, a_{k+1}\}$, with $Corr(A') \geq \gamma$, since the conditions of Theorem 1 may not hold due to a newly added item $a_{k+1}$. However if *all* $(h, k\text{-}1)$-itemsets in cell $Q_{h,k}$ are non-positive, then they cannot be combined in a positive itemset in cell $Q_{h,k+1}$.

Now, suppose that all $(h, k)$-itemsets in cell $Q_{h,k}$ and all $(h+1, k)$-itemsets in cell $Q_{h+1,k}$ are non-positive. Then, according to the flipping-based pruning we can terminate only the vertical extension to the next abstraction level. The following theorem proves that there are no flipping correlation patterns also to the right of column $k$, and thus we can terminate both the vertical and the horizontal pattern growth.

THEOREM 3. *Termination of the pattern growth (TPG)* *If all itemsets in $Q_{h,k}$ and $Q_{h+1,k}$ are non-positive, there are no flipping patterns in any column $k'$ for $k' \geq k$.*

PROOF. For every parent itemset $A_p$ in $Q_{h,k+1}$ and child itemset $A_c$ in $Q_{h+1,k+1}$, we know that they are non-positive by Corollary 1. By induction, we conclude that any $(h, k')$-itemset and $(h+1,k')$-itemset which are supersets of $A_p$ and $A_c$ respectively ($k' > k+1$) are non-positive. Therefore, any $k'$-itemset with $k' \geq k$ is not a part of a flipping pattern. ■

Finding that all itemsets in two subsequent cells in the same column are non-positive allows us to terminate the pattern growth. For example, if all patterns of $Q_{1,3}$ and $Q_{2,3}$ are negative as illustrated in Figure 7(a), then, based on Theorem 3 (TPG principle), we do not need to explore any cell below and to the right of cell $Q_{2,3}$.

In order to be able to check the termination condition at each step of the algorithm, we need to have at hand the results for two consecutive cells $Q_{h,k}$ and $Q_{h+1,k}$ in column $k$. Hence, the row-wise processing is adjusted. We first compute two upper rows of the search space table by zigzag, as illustrated in Figure 7(b): $Q_{1,2} \rightarrow Q_{2,2} \rightarrow Q_{1,3} \rightarrow Q_{2,3} \rightarrow \cdots \rightarrow Q_{1,K} \rightarrow Q_{2,K}$ until either the TPG termination condition is satisfied, or all itemsets in some cell are infrequent. Then, we process the remaining rows, one at a time. This ensures that we always have two cells in subsequent hierarchy levels, to apply the termination principle.

### 4.3.2 *Pruning single items and their supersets*

Corollary 2 suggests the pruning method for all itemsets containing a single item $a$, which satisfies the following conditions: (1) item $a$ has the smallest support between single items existing in the database, and (2) all $k$-itemsets containing $a$ are non-positive. This pruning is performed as follows.

Let $\mathcal{I}_h$ be a complete set of items at abstraction level $h$. The items from each $\mathcal{I}_h$ are sorted by support and are kept in list $\mathcal{L}_h$. Now, while computing $(h, k)$-itemsets in cell $Q_{h,k}$, for each item $a_i$ in $\mathcal{L}_h$ we keep track of the maximum *Corr* value among $(h, k)$-itemsets containing $a_i$. As a result, if we have that item $a_1$ with the smallest support on the top of $\mathcal{L}_h$ has maximum *Corr* below $\gamma$, we conclude, by Corollary 2, that all supersets of $a_1$ in subsequent columns of the search space table starting from $k + 1$ are non-positive. If we were to compute just positive correlations, we would implicitly remove item $a_1$ from the database. After removing it, another item, $a_2$ becomes the item with the smallest support, and if the above condition holds for item $a_2$, we could remove it too. We could continue removing items from the top of $\mathcal{L}_h$, until, for some item $a_j$, a positively correlated $(h, k)$-itemset exists. Now we have a set $\mathcal{R}_h$ of $j - 1$ items, for which we know that all their supersets of size more than $k$ are non-positive, and which present candidates for removal from the database.

After computing $k$-itemsets in at least two consecutive cells $Q_{h,k}$ and $Q_{h+1,k}$, we have two lists $\mathcal{R}_h$ and $\mathcal{R}_{h+1}$ of single items, whose supersets of size more than $k$ are non-positive. Then, for each item $a_i$ from $\mathcal{R}_{h+1}$, if its higher-level abstraction is in $\mathcal{R}_h$, then all supersets of $a_i$ are not a part of a flipping pattern, and thus, can be pruned.

We call this pruning method *Single-Item Based Pruning* (SIBP).

## 4.4 The complete algorithm

Algorithm 1 presents the pseudocode of Flipper. At lines 2-7, we compute two ceiling rows of the search space, counting itemsets in both cells for each $k$ simultaneously. We apply the TPG principle to terminate the horizontal extension as early as possible. At lines 8-15, we compute the rest of the search space in the row-wise manner. All $k$-itemsets which contain single items disqualified by the SIBP principle are pruned. Also, we apply the termination condition TPG to check whether we can terminate the horizontal extension.

**Algorithm 1**: The Flipper Algorithm. TPG stands for termination of the pattern growth (Theorem 3), and SIBP stands for pruning based on a single item (Theorem 2 and Corollary 2).

**input** : a transactional database $\mathcal{D} = \{D_1, D_2, ..., D_n\}$, a *taxonomy tree* $\mathcal{T}$, correlation thresholds $\gamma$ and $\varepsilon$, minimum support thresholds $\theta_h$ for $0 \leq h \leq H$
**output** : all flipping patterns

1 scan $\mathcal{D}$ and find frequent 1-items for each taxonomy level;
2 **for** $k = 2, \cdots, K$ **do**
3     scan $\mathcal{D}$ to compute *Corr* for all candidates $(1, k)$-itemsets and $(2, k)$-itemsets;
4     prune based on support, flipping and SIBP;
5     **if** *TPG*$(Q_{1,k}, Q_{2,k})$ **then** break;
6 **end**
7 eliminate non-flipping patterns in rows 1 and 2;
8 **for** $h = 3, \cdots, H$ **do**
9     **for** $k = 2, \cdots, K$ **do**
10         scan $\mathcal{D}$ to compute *Corr* measure for candidate itemsets in $Q_{h,k}$;
11         prune based on support, flipping and SIBP;
12         **if** *TPG*$(Q_{h-1,k}, Q_{h,k})$ **then** break;
13     **end**
14     eliminate non-flipping patterns in rows $h$-1 and $h$;
15 **end**
16 Check each non-empty $Q_{H,k}$ and report flipping patterns.;

# 5. EXPERIMENTAL EVALUATION

We present evaluation of pruning principles described in the previous section. The goal of performance experiments was to see if the number of candidates to be evaluated drops significantly by using proposed pruning techniques, in addition to support-based pruning. To assess the pruning power of each principle we started from a baseline version - the level-wise Apriori algorithm ("BASIC"), and then incrementally enhanced it with pruning by flipping ("FLIPPING PRUNING"), termination of pattern growth ("TPG"), and single-item based pruning ("SIBP"). The BASIC Apriori algorithm can be regarded as the baseline and represents all previous methods, which were computing all frequent patterns before ranking the correlations by surprisingness [6], or before removing the redundancy [12]. We test our new methods in the Apriori-like framework due to the simplicity of modeling the search space as a two-dimensional table. In all experiments, we use *Kulc* correlation measure, which is more tolerant for finding correlations in unbalanced datasets [22].

All versions perform counting by sequential scans of disk-resident input data. Thus, in general, they scale to massive inputs. The main memory is used to store the remaining candidates. The candidates are pruned after finishing each cell $Q_{h,k}$, in order to keep the usage of the main memory to a possible minimum. The experiments were performed on a Linux (ver 2.6.18) server with quad core Xeon 5500 processors and 48 GB of main memory. The BASIC consumed up to 40GB of RAM to store all the candidates, while the enhanced versions never required more than 2GB of memory. In experiments with real datasets, we found that in order to produce flipping patterns we need to set the minimum support threshold very low, which did not allow us to compare with the "BASIC" Apriori algorithm. For such low

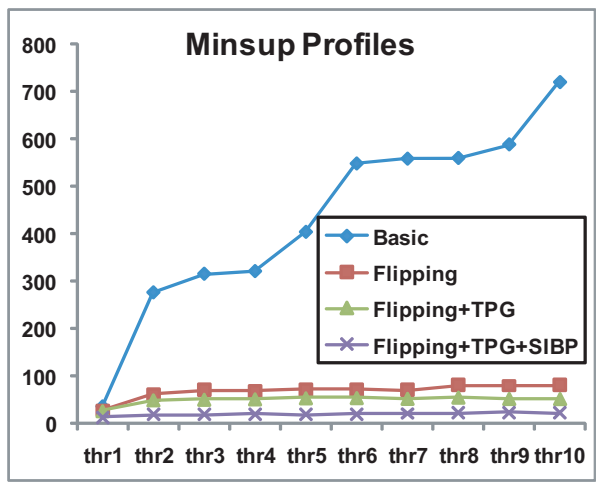

(a) Running time (sec) *w.r.t.* minimum support profile $(\theta_1, \theta_2, \theta_3, \theta_4)$

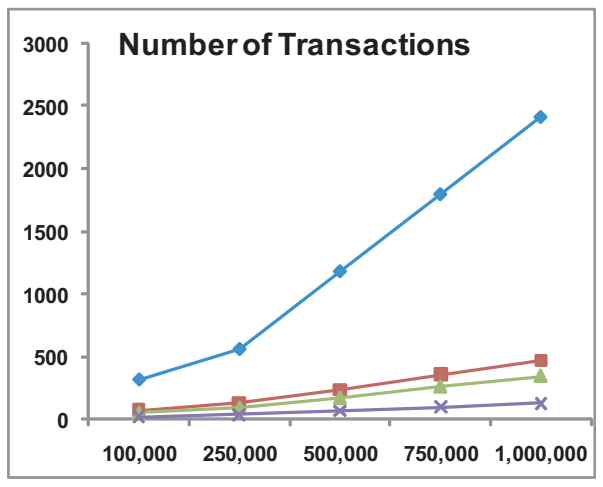

(b) Running time (sec) *w.r.t.* number of transactions

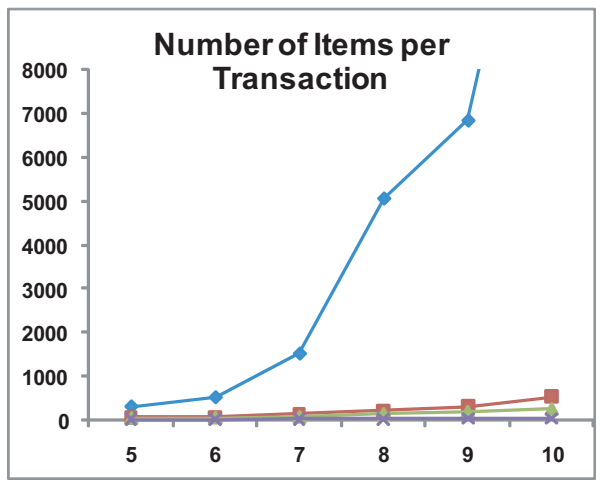

(c) Running time (sec) *w.r.t.* transaction width

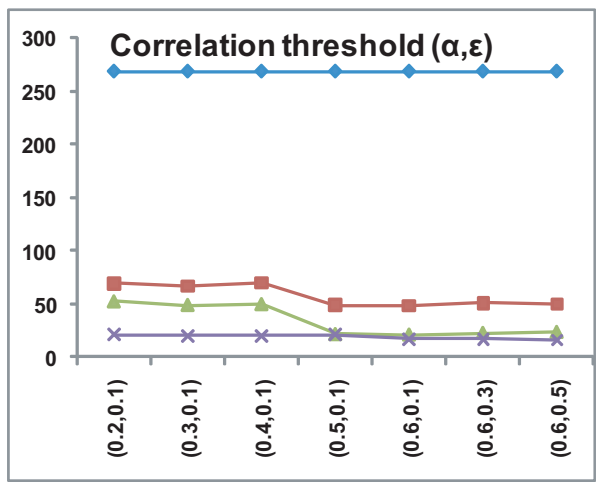

(d) Running time (sec) *w.r.t.* correlation thresholds

**Figure 8: Performance for synthetic datasets.**

supports, the exponential explosion of candidate itemsets to be kept simultaneously in main memory leads to memory overflow and disk thrashing. In contrast, the number of remaining candidates after pruning by new methods is reasonably small (see Table 4). This demonstrates that, for a low-support range, FLIPPER is significantly more scalable than existing support-pruning based mining algorithms. In addition, note that in our approach we generate a small subset of unexpected patterns, rather than the complete pool of frequent itemsets.

## 5.1 Synthetic datasets

We studied the influence of different parameters on the performance of FLIPPER. For this, the synthetic datasets were generated using generator by Srikant and Agrawal [17]. We have set the following default parameter values: number of transactions $N = 100K$, average number of items per transaction (transaction width) $W = 5$, number of distinct items $|\mathcal{I}| = 1,000$, number of hierarchy levels $H = 4$. The number of distinct categories at the first level is 10, the fanout is 5. The default set of thresholds is as follows: minimum support thresholds ($\theta_1 = 1\%, \theta_2 = 0.1\%, \theta_3 = 0.05\%, \theta_4 = 0.01\%$) and correlation thresholds ($\gamma = 0.3, \varepsilon = 0.1$).

**Minimum Support:** Because we used 4 minimum support thresholds, one for each level of the hierarchy, we made a value-decreasing sequence of 10 minimum support threshold profiles described in Table 3. Profile thr1 is a profile with high support thresholds for all levels. Starting from thr2, we lowered minimum support thresholds for each hierarchy level one at a time.

The results are shown in Figure 8(a). For the case of a high minimum support (thr1), the running time is low for all methods, indicating that pruning based only on support (BASIC) works well for high minimum support thresholds. However, for lower minimum supports pruning by support becomes insufficient. The minimum support threshold at the bottom level of hierarchy $\theta_4$ has the largest impact on the performance. We observe a sudden increase in the running time of our baseline method for thr2, thr6 and thr10, when $\theta_4$ is lowered. This is because the largest number of distinct items is on the bottom level of the hierarchy. For low minimum supports, the total number of frequent patterns explodes. Using all of the new pruning techniques together makes the computation up to 30 times faster.

**Table 3: Minimum support profiles**

| Profile | $\theta_1$ | $\theta_2$ | $\theta_3$ | $\theta_4$ |
|---|---|---|---|---|
| thr1 | 0.05 | 0.05 | 0.05 | 0.05 |
| thr2 | 0.05 | 0.001 | 0.0005 | 0.0001 |
| thr3 | 0.01 | 0.001 | 0.0005 | 0.0001 |
| thr4 | 0.01 | 0.0005 | 0.0005 | 0.0001 |
| thr5 | 0.01 | 0.0005 | 0.0001 | 0.0001 |
| thr6 | 0.01 | 0.0005 | 0.0001 | 0.00005 |
| thr7 | 0.001 | 0.0005 | 0.0001 | 0.00005 |
| thr8 | 0.001 | 0.0001 | 0.0001 | 0.00005 |
| thr9 | 0.001 | 0.0001 | 0.00006 | 0.00005 |
| thr10 | 0.001 | 0.0001 | 0.00006 | 0.00003 |

**Number of Transactions:** In Figure 8(b), we used 5 different datasets varying $N$ from 100K to 1M. For all methods, the running time shows linear dependency on $N$. With all new pruning methods, FLIPPER runs 15–20 times faster than the baseline method.

**Average transaction width:** Figure 8(c) shows results for 6 different datasets with default parameters, where the average transaction width $W$ is increased from 5 to 10. By increasing $W$ we get more frequent patterns. For larger $W$ we see a dramatic increase in running time for our baseline method, while our new techniques handle the increasing density gracefully. FLIPPER with full pruning could run up to 5, 10, and 300 times faster than FLIPPING+TPG, FLIPPING, and BASIC methods respectively.

**Correlation Thresholds:** Because we have two parameters ($\gamma$, $\varepsilon$) for correlation thresholds, we used the value-increasing sequence of 7 profiles for this experiment. For the first 5 profiles we fixed the negative threshold $\varepsilon$ as 0.1 and increased positive threshold values by 0.1, and for the rest we fixed the positive threshold $\gamma$ as 0.6 and increased negative threshold values by 0.2.

We remind the reader that our advanced pruning is based on a non-positivity of candidate patterns. Hence, the efficiency of pruning grows when $\gamma$ becomes larger and the number of positive itemsets drops. Figure 8(d) shows the corresponding result: the larger is $\gamma$, the more candidates are pruned by all 3 pruning methods, and the faster is the computation. Note that the baseline method does not depend on correlation thresholds, since it generates all frequent itemsets and disregards the correlation values.

The general conclusion from these experiments is that if we want to obtain correlations in itemsets with *low supports* in *dense* transactional databases, using the baseline Apriori algorithm is computationally infeasible, and the new pruning methods are quite useful for this scenario.

Based on this performance evaluation, we may suggest the following guidance for parameter settings. First, different support thresholds should be set for each level of the hierarchy. The best strategy is to set support thresholds comparatively high at the upper levels, and then lower them to the more detailed level. The support for the bottom level should be set considerably low, otherwise all the itemsets could be pruned. Such low level of support was unattainable by previous methods, due to the enormous number of candidates which need to be considered. Second, the data expert should set the positive correlation threshold $\gamma$. The efficiency of FLIPPER is due to the pruning of non-positive itemsets, so the main performance factor is the careful choice of $\gamma$. Then the user may start from setting the negative threshold just below $\gamma$, and gradually decrease it until the satisfactory number of flipping patterns is obtained.

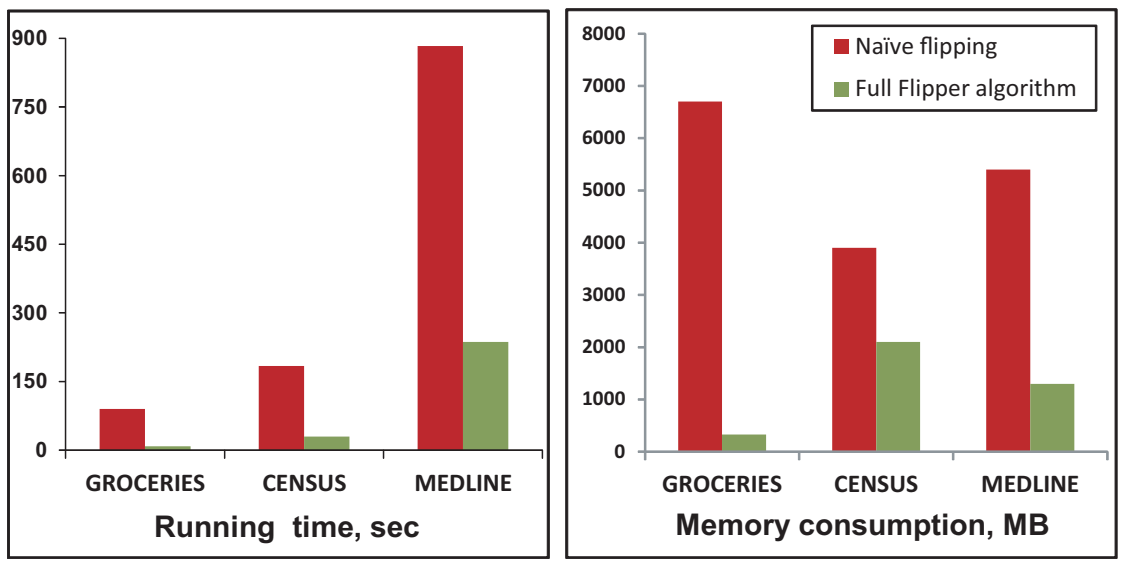

(a) (b)

**Figure 9: Comparative performance for real datasets.**

## 5.2 Reality check

We applied the market-basket concept to the following real-life datasets:

The **GROCERIES** dataset [5][2] represents 1-month of the point-of-sale transactions in a local grocery store. The taxonomy of items is provided and it represents item categorization used in this store. The dataset contains 9,800 transactions, and the taxonomy has three levels of abstraction.

The **CENSUS** dataset [10][3] is an extract from the US Census 1996. It represents multi-attribute records, where each record characterizes a single person. Income attribute is discretized into two bins: income $\geq \$50K/yr$ or $< \$50K/yr$. We considered each record as a transaction. We manually created hierarchies with two and three levels based on different attribute combinations. For example, for the group of people *occupation:executive and sex: women*, the higher level generalization is all people with *occupation:executive*. Then the flipping pattern would be to find that *occupation:executive* is strongly positively correlated with income $\geq \$50K/yr$ and that this correlation becomes negative for the sub-population women executives. This dataset contains 32,000 transactions.

The **MEDLINE** dataset is a set of the medical paper citations. Each citation (paper) is a transaction. The items are the topics. The hierarchy of topics was obtained from the Medical Subject Headings database (MeSH)[4]. This hierarchical terminology was used to manually index each article in MEDLINE database. Each paper contains one or several categorical topics assigned to it. Our working set contains all medical papers published in year 2010[5] (640,000 citations), and we consider only three top levels of the detailed hierarchy tree[6].

In Figure 9(a), we show the performance results for the naïve flipping-based pruning vs. full FLIPPER with our two new pruning techniques. We exclude the baseline Apriori method, which runs longer than 10 hours even for the smallest dataset GROCERIES. Figure 9(b) shows the memory consumption by these two programs. Again, the baseline Apriori method is not included because it requires more than 48 GB to store all frequent candidates, more memory than was available to us. Note that the space was used to store *all* candidate itemsets with their counts, and still our full version never required more than 2 GB of RAM. This points out that mining correlations directly produces a small subset of candidates, which can be efficiently handled by a modern machine. It is possible to further reduce memory consumption by exploring various optimization methods: for example, for generating candidates in cell $Q_{h,k}$ we only need to keep the results for $Q_{h,k-1}$, and to apply the termination of the pattern growth and the single-item based pruning we need to simultaneously keep in memory the results only for two consecutive cells $Q_{h-1,k}$ and $Q_{h,k}$. The candidates from the other cells can be eliminated, making the number of itemsets in memory reasonably small. This confirms the efficiency of our method which generates a small number of interesting patterns, without enumerating a large amount of candidates.

The flipping patterns were almost absent from synthetic datasets used in our experiments, but for real datasets, we could produce a reasonable amount of flipping patterns as shown in Table 4. Note that for the low-support profiles used in our experiments, the number of flipping patterns is substantially lower than the total number of all negative and positive patterns. For high support thresholds, the total number of positive and negative patterns decreases significantly, but all these patterns are trivial, and none of them is flipping. On the other hand, many out of the discovered flipping patterns are interesting and unexpected. Of course, they are contained in the set of all positive and negative patterns, however it is much harder to find them there. Moreover, for larger datasets the computation of the entire set of all negative patterns is infeasible.

**Table 4: Number of flipping patterns vs. all positive and negative frequent patterns in datasets GROCERIES(G), CENSUS(C) and MEDLINE(M).**

| | $(\gamma, \varepsilon, \theta_1, \theta_2, \theta_3)$ | Pos | Neg | Flips |
|---|---|---|---|---|
| G | (0.15, 0.10, 0.001, 0.0005, 0.0002) | $4.8 \cdot 10^3$ | $8.0 \cdot 10^4$ | 174 |
| C | (0.25, 0.15, 0.002, 0.001, 0.0001) | $1.4 \cdot 10^5$ | $7.3 \cdot 10^4$ | 232 |
| M | (0.40, 0.10, 0.001, 0.0005, 0.0001) | $4.2 \cdot 10^3$ | $1.6 \cdot 10^6$ | 430 |

In Figures 10,11 and 12 we present a pair of the flipping correlation patterns for each dataset. Each example shows positive or negative correlation between a pair of items at the most detailed level of abstraction (bottom), accompanied by the corresponding contrasting correlations between their abstractions at higher levels.

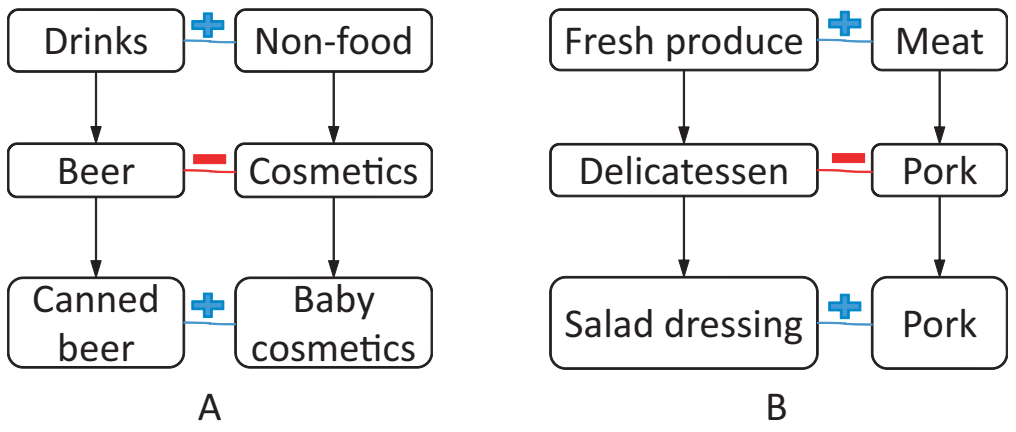

**Figure 10: Flipping correlation in Groceries dataset.**

**Flipping in GROCERIES dataset** (Figure 10). **Pattern A** reflects a famous itemsets $\{beer, diaper\}$, now in a more highlighted way: by showing the negative correlation between their minimal generalizations. The second example demonstrates that the flipping patterns can be used to design more user-friendly store layouts. It happens often that customers expect to find some product combinations in close proximity while by store design these items belong to different and unrelated categories. For example, in **Pattern B**, pork and salad dressing are positively correlated, while in general pork and delicatessen are negatively correlated. This might suggest removing the salad dressing from delicatessen, and moving it closer to the meat department. Many other patterns from this dataset are *surprising* and *actionable*. For example, the strong negative correlation between eggs and fish is accompanied by positive correlation between their higher categories, fresh products and meat&fish. The strong positive correlation between baby cosmetics and oil is highlighted by the negative correlation of such unrelated product categories as cosmetics and oils.

[2]http://rss.acs.unt.edu/Rdoc/library/arules/data/
[3]http://archive.ics.uci.edu/ml/datasets/Adult
[4]http://www.nlm.nih.gov/mesh/
[5]http://mbr.nlm.nih.gov/Download/index.shtml
[6]http://www.nlm.nih.gov/mesh/2010/mesh_browser

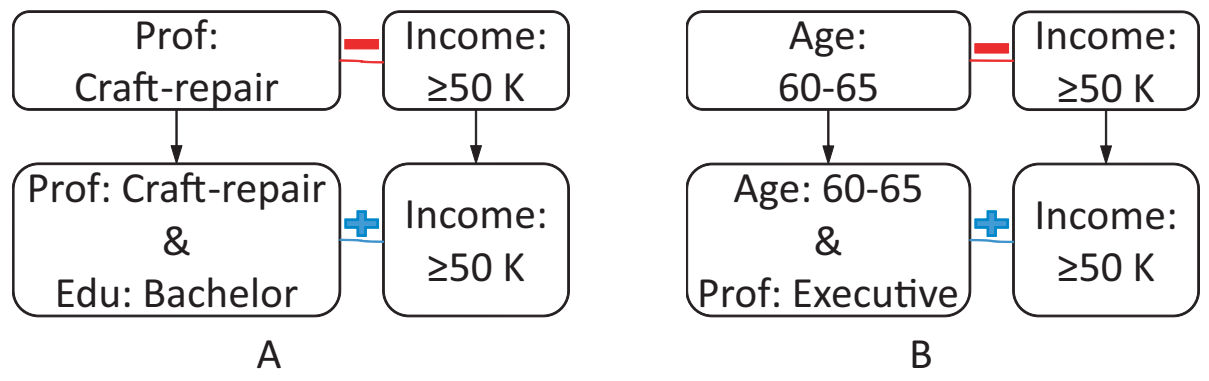

**Figure 11: Flipping correlation in Census dataset.**

**Flipping in CENSUS dataset** (Figure 11). These patterns suggest that Flipper can be used to compare characteristics of different sub-populations organized into hierarchical categories. From **Pattern A** we learn that education matters: people working in Craft-repair and having Bachelor degree are positively correlated with income $\geq \$50K/yr$, while their generalization group – all people working in Craft-repair are negatively correlated with $\geq \$50K/yr$. **Pattern B** suggests that it is hard to get 50 K per year if you are at age group 60–65, unless you are an executive.

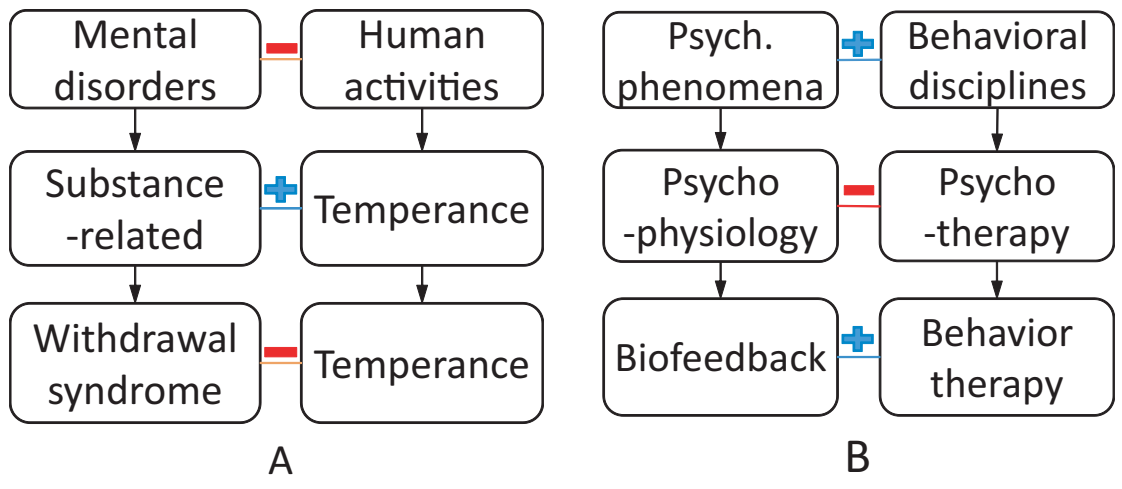

**Figure 12: Flipping correlations in MEDLINE dataset.**

**Flipping in MEDLINE dataset** (Figure 12). The suggestions of new research topic combinations obtained from this dataset can be used by researchers in medical field. **Pattern A**: if substance-related mental disorders were often studied together with temperance, then it is quite reasonable to research the combination of the withdrawal syndrome with temperance, underrepresented in current medical publications. **Pattern B** may suggest the collaboration between two unrelated areas of psychophysiology and psychotherapy. However, if one decides to study the combination of such sub-topics as biofeedback and behavior therapy, he finds out that these two are often studied together.

To summarize, flipping correlation patterns can be used to find items which were incorrectly assigned to the wrong category; to find surprising non-trivial correlations to be explained; to discover underrepresented, or overrepresented combinations of items; or to discover correlations specific for some sub-population. All these possible new insights into the data become possible with our new approach.

# 6. RELATED WORK

The market basket concept [1] was generalized into a notion of correlations in a pioneering work of Brin et al. [3]. The common approach is to compute correlations between items in each frequent itemset, implying that all frequent itemsets should be generated first. The best pattern mining algorithms (e.g., [1, 8]) rely heavily on the suppot-based pruning. However, *low* support thresholds have an adversary effect on the efficiency of these algorithms, calling essentially for counting all possible combinations of items and leading to a very inefficient (and often infeasible) computation. This is especially true when the goal is to include negative correlations, which by definition occur in itemsets with very low support ([3, 15, 16, 23, 2, 22]). And so the complete set of all negatively correlated items is so far impossible to produce directly [20]. An indirect method for computing negative correlations [15] is called "support expectation based on concept hierarchy". Similar to other expectation-based techniques, the correlations estimated by this method cannot be reliable because the measure is not null-invariant.

The focus of the research has shifted from producing a complete set of all correlations to discovering only interesting and non-trivial patterns among the vast number of all possible frequent patterns. Multiple interestingness measures were proposed, both subjective [13] and objective [19]. A comprehensive review of different interestingness measures can be found in the book by Hilderman and Hamilton [9]. The idea of using taxonomies for pruning of redundant correlations (rules) was first introduced by Srikant and Agrawal [17]. Ranking correlations (rules) based on the distance between participating items in a given taxonomy tree was studied by Hamani and Maamri [6]. This is a particular example of a more general approach [13], where a rule (positive correlation) is considered interesting, only if it contradicts to a rule from a set of pre-defined user beliefs. In this example [6], the user beliefs are presented in a form of hierarchical categories, and all the high-confidence rules are ranked by surprisingness, which is proportional to the number of edges on the shortest path between taxonomy tree nodes (items). A similar work [7] discusses how to mine certain "level-crossing" rules. [14] extends the previous works to mine multilevel association rules directly from hierarchies.

To our knowledge, the mining and use of flipping correlation patterns as highly contrasting level-specific correlations has not been proposed or studied before. The use of null-invariant correlation measures was discussed recently in the context of positive correlations [22]. We extend these measures to negative correlations. Though the pruning based on a non-antimonotonic null-invariant correlation measure ($Kulc$) was introduced in work of Wu et al. [22], the proposed pruning turned out to be efficent only for high supports and for high correlation thresholds, and thus could not be used as a baseline for mining flipping patterns, which includes finding correlation *below* the negative correlation threshold.

Our mining methodology is novel as well since unlike previous works (such as [6]) we directly mine the flipping correlations thereby eliminating the inefficient frequent itemsets mining step. Our results clearly indicate that previous approaches, represented by our baseline experiments, are not as efficient nor as expressive as the flipping correlation mining method introduced in this paper.

# 7. CONCLUSIONS AND FUTURE WORK

In this paper, we introduced a new concept of flipping correlation patterns. We presented an efficient algorithm, FLIPPER, for mining these new patterns. Despite the fact that the selected correlation measure, *Kulc*, is not anti-monotonic, we developed sharp pruning techniques, based on flipping constraints and mathematical properties of *Kulc*. We generalized these new techniques to all popular null-invariant correlation measures, such as *Coherence*, *Cosine*, *All Confidence* and *Max Confidence*. In our experiments with low support thresholds, we demonstrated a high efficiency of a new correlation-based pruning compared to the pruning based exclusively on support. Using real datasets, we have shown that interesting new observations can be extracted from the data by using the flipping pattern concept.

As future work, the flipping pattern concept can be extended for discovering a set of discriminative correlations, that are specific for a given sub-group. Another challenging topic for future research is the choice of threshold values $\gamma$ and $\varepsilon$. This is because a data expert might not be able to say which correlation value should be considered positive or negative in a particular dataset. One of the possible solutions to this problem is to produce top-$K$ "most flipping" patterns, which could be defined as the patterns with the largest gap between correlation values at different hierarchy levels. FLIPPER is the first algorithm for level-specific contrasting correlations. It uses new correlation-based pruning methods in a simple Apriori-like framework. The use of more advanced data structures and more advanced pruning methods might also be an area of a fruitful future research.

## Acknowledgement

The work was supported in part by the Canadian Postdoctoral NSERC Fellowship *PDF*-388215, the U.S. National Science Foundation grant *IIS*-09-05215, and the NDSEG fellowship for the third author. Any opinions, findings, and conclusions expressed here are those of the authors and do not necessarily reflect the views of the funding agencies.